\documentstyle[aps,prb,epsf,floats]{revtex}

\newcommand{\be}{\begin{equation}}
\newcommand{\ee}{\end{equation}}
\newcommand{\bi}{\bibitem}
\newcommand{\D}{\Delta}
\newcommand{\chiSG}{\chi_{\scriptscriptstyle SG}}
\newcommand{\rDT}{r_{\scriptscriptstyle \Delta T}}
\newcommand{\rDJ}{r_{\scriptscriptstyle \Delta J}}
\begin{document}
\twocolumn[\hsize\textwidth\columnwidth\hsize\csname@twocolumnfalse%
\endcsname
\title{
Chaos
in a Two-Dimensional Ising Spin Glass}
\author{Muriel Ney-Nifle\cite{perm:add} and A. Peter Young}
\address{Department of Physics, University of California\\
Santa Cruz CA 95064}
\maketitle
\begin{abstract}
We study chaos in a two dimensional Ising spin glass by finite temperature
Monte Carlo simulations. We are able to detect chaos with respect to
temperature changes as well as chaos with respect to changing the bonds, and
find that the chaos exponents for these two cases are equal. Our value for the
exponent appears to be consistent with that obtained in studies at zero
temperature.
\end{abstract}
%\pacs{PACS numbers: 71.45.Lr,72.70.+m,74.60.Ge}
\vskip 0.2 truein
]

\newpage
\section{Introduction}
\label{intro}

A characteristic feature of spin glasses is that the relative orientations of
the spins in
the spin glass state are not uniquely determined by the model, but rather vary
with external parameters such as the temperature or the magnetic field.  At
large separation, $R$, the correlation between the spins
varies in a chaotic manner as a function of temperature, and, when the
temperature is altered by an amount $\Delta T$, will change substantially at
distances $R$ greater than $l_{\D T}$ where
\begin{equation}
l_{\Delta T} \sim (\D T)^{-1/\zeta},
\label{zeta} 
\end{equation}
which defines the chaos exponent $\zeta$.   This temperature induced chaos has
been quite difficult to see in Monte Carlo (MC)
simulations\cite{r,kss} or mean field theory\cite{k,kv}, and claims have
been made that is absent or very small\cite{r,kss,k}.  A larger chaotic
effect has been observed when making a small change
%, $\Delta J$,
in the
couplings~\cite{afr,rsb}.

Chaos with temperature has been shown to be a common feature of the two main
models for the spin glasses phase: mean field theory \cite{k,kv} 
and the droplet
theory \cite{bm,fh,nh1,nh}. 
The later is based upon real space renormalization group
calculations which allow for quantitative results. In particular, they indicate
that a temperature perturbation generates a disorder perturbation and thus
these two perturbations should have the {\em same} chaos exponent \cite{nh}.

Here we study chaos with both $\Delta T$ and a change in the
couplings, $\D J$  (defined precisely in Eq.~(\ref{dJ}) below),
by Monte Carlo simulations
for a two-dimensional Ising spin glass at finite temperature.
Since $T_c=0$ for this model, all our data is in the paramagnetic phase, where
the correlation function tends to zero at large distances, so we are looking
for chaos in the sign of this decaying function (such chaos has been 
shown
in a one dimensional system \cite{nh1}).   We study distances
which are smaller than the correlation length so the chaos we obtain is that
corresponding to the critical point\cite{nh,footnote1}.

Our main results are as follows:
\begin{enumerate}
\item
It {\em is} possible to see chaos with $\D T$ as well as chaos with $\Delta J$.
\item
The chaos exponents for $\D T$ and $\Delta J$ appear to be equal.
\item
The chaos exponent found here at finite-$T$ seems to be consistent with that
obtained at $T=0$~\cite{rsb,bm},
%$\relbar$
%though, given the uncertainty due to
%systematic errors, we cannot be completely sure about this. 
%If the exponents really are different
%then there is a violation of the scaling theory in two dimension.
\end{enumerate}

The plan of the paper is as follows: In \S \ref{model} we define the model and
various quantities of interest. \S \ref{fss}
discusses finite-size effects which will
be very important for the analysis  while 
\S \ref{results} presents the numerical
results that are then interpreted in \S \ref{interpret}.
%Our conclusions are summarized
%in \S \ref{concl}.

\section{The Model}
\label{model}

We consider the Edwards-Anderson Hamiltonian with Ising spins and nearest
neighbor couplings,
\be
{\cal H} = -\sum_{\langle i, j\rangle} J_{ij} S_i S_j \ ,
\ee
where the $\lbrace J_{ij}\rbrace$ are drawn from a Gaussian distribution with
zero mean and variance $[J_{ij}^2]_{av}$ equal to unity. We denote by
$[\cdots]_{av}$ an average over the interactions. The spins lie on a square
lattice of linear size $L$ with periodic boundary conditions.

For each realization of the disorder we simulate several copies (or replicas)
of the system.  The basic quantity we calculate is the replica overlap 
\be
q_{ab} = {1 \over N} \sum_{i=1}^N S_i^{(a)} S_i^{(b)} \ ,
\label{qab}
\ee 
where $a$ and $b$ denote replicas and $N=L^2$.
When we investigate chaos with $\D T$, some of the replicas will
have identical bonds but slightly different temperatures and when we investigate
chaos with $\D J$ some of the replicas will have slightly different couplings
but the same temperature. 

Next we describe quantities that we calculated in the simulations. First of
all, from replicas with the {\em}
{\em same} temperatures and bonds, we compute the standard equilibrium
quantities,
$g$, the Binder ratio, and 
$\chiSG$, the spin glass susceptibility, defined by
\begin{eqnarray}
g  & \equiv &  {1 \over 2} 
%\left[ 3 -  {{\left[ \langle q^4 \rangle \right]_{av}} \over 
%{\left[ \langle q^2 \rangle^2 \right]_{av}}} \right] \label{g} \\
\left[ 3 - 
{ \langle q^4 \rangle \over \langle q^2 \rangle^2 }
\right]
\label{g} \\
\chiSG & \equiv &
L^2  \langle q^2 \rangle \ ,
\label{chi}
\end{eqnarray}
where 
$\langle \cdots \rangle$, denotes both the average over disorder and the
statistical mechanics (Monte Carlo) average.
During the simulation,
the first $t_0$ sweeps are used for equilibration and the next $t_0$
sweeps are used for measurements.  We check  that the system is in
equilibrium by standard methods\cite{by}.
The equilibration time $t_0$ limits the maximum size and minimum temperature we
can study. In our case, we can reach $T = 0.4$ for $L = 6$ and $T =
0.55$ for $L = 10$, which both require about $t_0 = 10^6$ MC steps.
%Our faster computers do $2.10^6$ up-dates per second. 

Next we describe quantities that we calculate to determine the chaos. First,
%We now come to the study of the chaotic behavior of the correlation function.
consider chaos with changing the bonds, keeping the
temperature fixed. This is done by running one replica with
a set of bonds $\lbrace J_{ij} \rbrace$ and another with bonds $\lbrace
J^\prime_{ij}\rbrace$, where\cite{nh}
\be
\label{dJ}
J^\prime_{ij} = {{J_{ij} + x_{ij} \D J}\over {\sqrt{1+{\Delta J}^2}}} ,
\ee
where $x_{ij}$ is a Gaussian random variable with 
zero mean and unit variance. Note that the
$\lbrace J^\prime_{ij}\rbrace$,
and the $\lbrace J_{ij}\rbrace$ have the {\em same} distribution. 
A convenient measure of how much the change in the bonds alters the spin
orientations is the dimensionless ``chaos parameter''\cite{afr},
\be
\rDJ \equiv 
{\langle q_{\scriptscriptstyle JJ^\prime}^2 \rangle \over \langle
q_{\scriptscriptstyle JJ}^2 \rangle} ,
\label{rJ}
\ee
%when $L < \xi$.
where the labels on the replicas refer to the bond distributions that are used.

When the temperature is changed
we consider
the overlap from replicas at temperatures {\em symmetrically} displaced
about $T$ as follows:
\be
\rDT \equiv 
{ \langle q^2_{\scriptscriptstyle T_+ T_-} \rangle \over
\sqrt{ \langle q^2_{\scriptscriptstyle T_+ T_+} \rangle \langle
q^2_{\scriptscriptstyle T_- T_-} \rangle} } ,
\label{rT}
\ee
where the temperatures are $T_{\pm} = T \pm \D T / 2$. 
We believe that this is the first time that chaos with temperature
has been calculated in this way. Other
attempts \cite{r} to look for chaos with temperature used a quantity
which has 
an asymmetry in temperature and will therefore involve
bigger corrections to scaling when the ratio $\D T / T$ is not vanishingly
small.

We average over from 80 to 400 realizations of the disorder. Error bars are
determined by grouping the results for the different samples into bins and
calculating the standard deviations among bins\cite{note}. 

\section{Finite Size Effects}
\label{fss}

To be in the scaling regime
it is necessary to work at moderately low
temperatures where finite size effects are important, and 
so finite-size
scaling\cite{privman} techniques are needed.

Since $\xi \sim T^{-\nu}$,
where $\xi$ is the {\em bulk} correlation length,
and since, at the $T=0$ critical point, the ground
state is unique, finite-size scaling predicts the following behavior for the
Binder ratio and $\chiSG$:
\begin{eqnarray}
g & = &  \tilde{g} \left(L^{1/\nu}T \right) 
\label{gscale} \\
\chiSG  & = &  L^2 \tilde{\chi}_{\scriptscriptstyle SG}
\left(L^{1/\nu}T \right) .
\label{chiscale}
\end{eqnarray}

%For our results on chaos we study a range of distance, $R$,
%such that
%$R < \xi$ (so, as discussed in \S \ref{intro}, the chaos is that of the
%critical point).
%Since the chaos parameters $r_{\D T}$ and $r_{\D J}$ sum over all distances, 
%this means that we need $L < \xi$, and so again finite-size scaling is 
%required.
%In the limit where $\xi$ is much larger than $L$, the dependence on $\xi$
%should drop out and one gets
We also need the finite size scaling ansatzes
for the chaos parameters, $\rDJ$ and
$\rDT$:
\begin{eqnarray}
\rDJ & = & \tilde{r}_{\scriptscriptstyle \D J}\left( L^\zeta \D J,
L^{1/\nu} T \right)
\label{rJscale} \\
\rDT & = & \tilde{r}_{\scriptscriptstyle \D T}\left( L^\zeta \D T ,
L^{1/\nu} T \right) ,
\label{rTscale} 
\end{eqnarray}
where we have used Eq.~(\ref{zeta}). It is inconvenient to analyze a function
of two variables, so, following a suggestion of D. Huse, we have taken data
where the second argument $L^{1\over\nu} T$, is roughly constant. We then
try to collapse the data on to a single curve by plotting it against $ L^\zeta
\D J$ with a suitable choice of $\zeta$. 
%where the dependence of $l_{\D T}$ of $\D T$ is given by Eq.~(\ref{zeta})
%and $l_{\D J}$ varies in a similar manner with $\D J$.

\section{Results}
\label{results}

\begin{figure}
\epsfxsize=\columnwidth\epsfbox{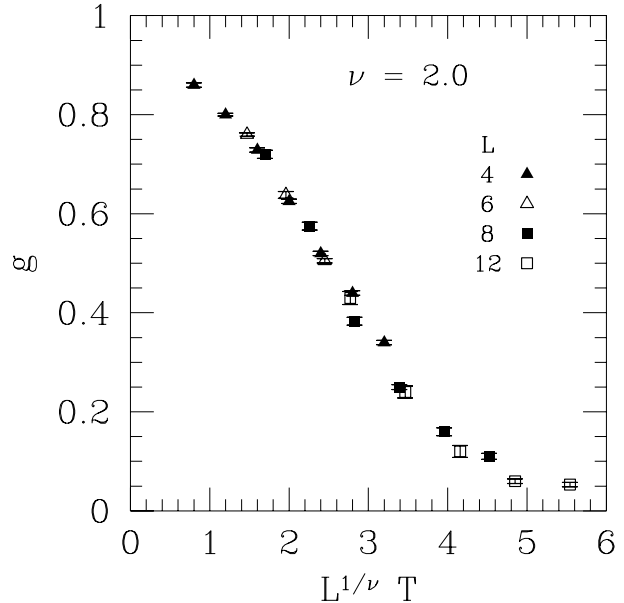}
\caption{
A scaling plot of the Binder ratio according to Eq.~(\protect\ref{gscale})
for different sizes and temperatures. The
value of the correlation length exponent is $\nu = 2.0$.
}
\label{plot:g}
\end{figure}

\begin{figure}
\epsfxsize=\columnwidth\epsfbox{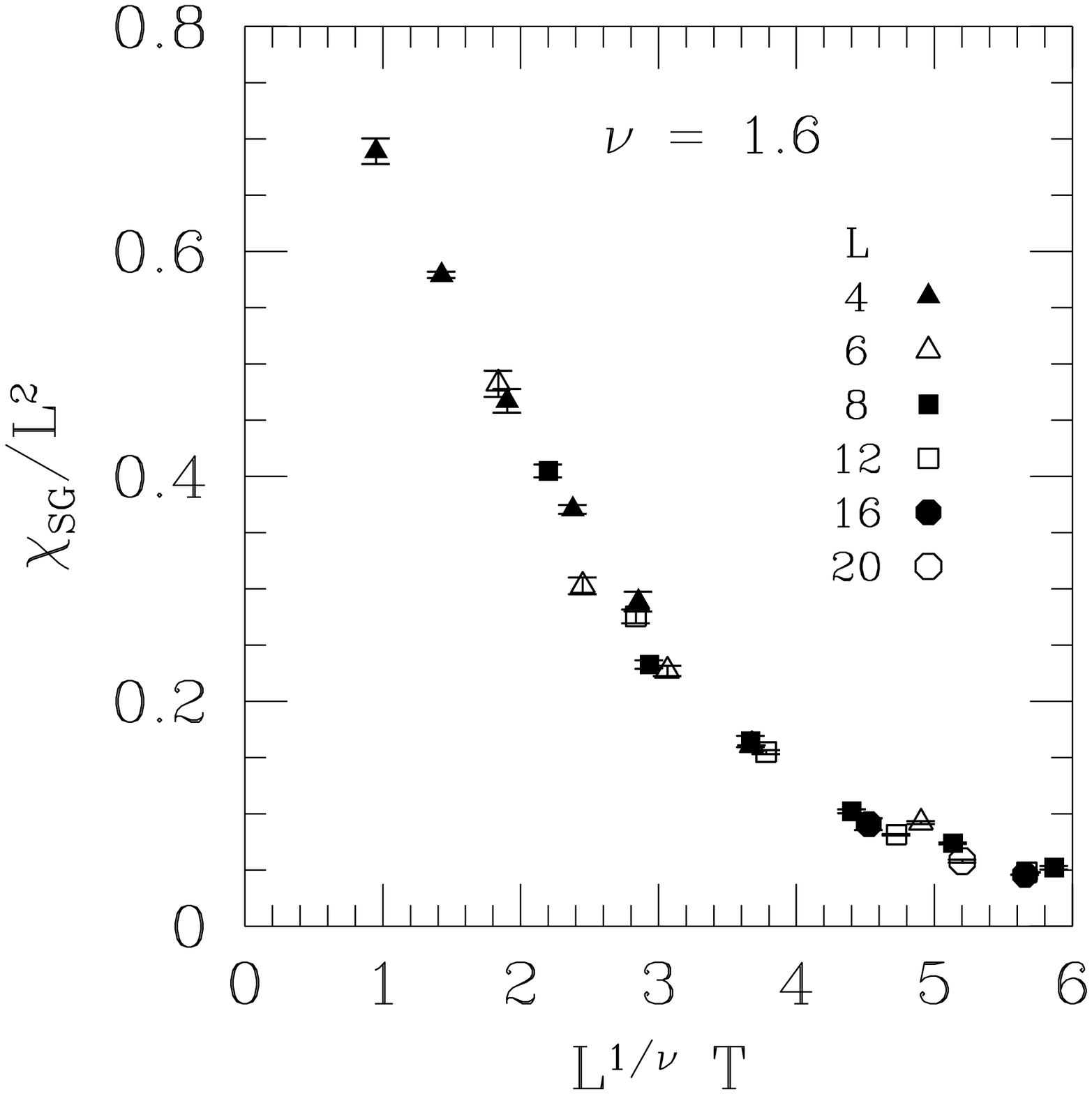}
\caption{
A scaling plot of $\chiSG$ according to Eq.~(\protect\ref{chiscale})
for different sizes and temperatures. The
value of the correlation length exponent is $\nu = 1.6$.
}
\label{plot:chi}
\end{figure}

Scaling plots for the Binder ratio and spin glass susceptibility are shown
in Figs. \ref{plot:g} and \ref{plot:chi}.
In Fig. \ref{plot:g} the sizes $L$
range from $4$ to $12$ and temperatures $T$ from $0.4$ to $1.6$.
The resulting correlation length exponent at the $T=0$ transition is
$\nu = 2.0 \pm 0.2$.
For Fig. \ref{plot:chi} the sizes extend from $L=4$ to $L=20$, at temperatures
from 0.4 to 1.6, and the correlation length
exponent is given by $\nu = 1.6 \pm 0.2$.

The value for $\nu$ obtained from $g$ is in good agreement with finite
temperature Monte Carlo simulations of Liang\cite{liang} and also agrees with
the work of Kawashima et al.\cite{khs}. The estimate for $\nu$ obtained from
$\chiSG$ is somewhat smaller, presumably reflecting the systematic corrections
to finite size scaling at the temperatures and sizes that we can study.
Interestingly the same trend, namely a larger value for $\nu$ obtained from $g$
than from $\chiSG$, has been seen in other models\cite{ky}.

Combining the two values for $\nu$ we estimate
\be
{\bar \nu} \ = \ 1.8 \pm 0.4 \ .
\label{nu}
\ee

We also determined the spin-spin correlation function for temperatures between
0.8 and 1.2. By fitting this data to an exponential function of position,
we estimate the correlation length, finding that
it could be fitted to $\xi = AT^{-\nu}$
with $A=4\pm 0.5$ and $\nu = 1.8\pm 0.2$,
the latter being consistent with the estimates from the finite-size scaling
analysis above.

%The goal here, though, is to also
%determine the prefactor, $A$, since
%we need to know, quite precisely,
%whether the size of the system is larger or smaller than the
%correlation length. Extrapolating the fit to $T=0.6$ (where finite-size effects
%preclude a direct determination of $\xi$), we find $\xi(0.6) =10\pm 2$. The
%scaling forms, Eqs.~(\ref{rJscale}) and (\ref{rTscale}), 
%for $\rDJ$ and $\rDT$ require that $\xi >
%L$ so we took $L \le 10$ at $T=0.6$ and $L \le 6$ for $T=0.8$. 

\begin{figure}
\epsfxsize=\columnwidth\epsfbox{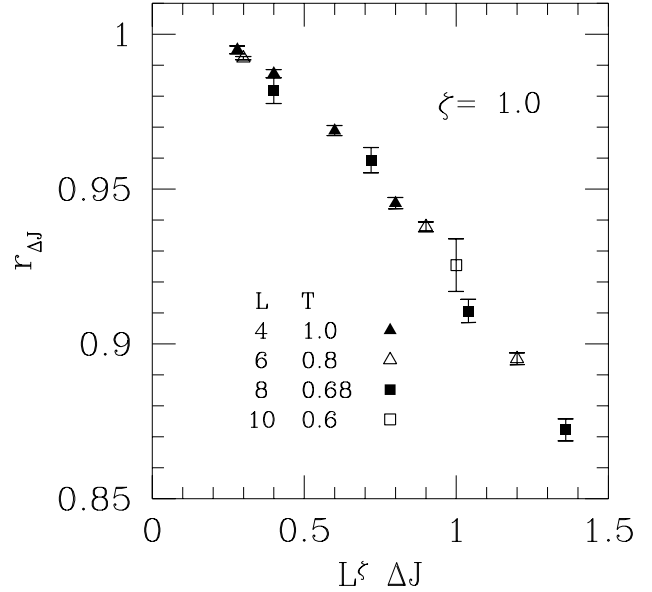}
\caption{
A scaling plot of $r_{\scriptscriptstyle \D J}$ according to
Eq.~(\protect\ref{rJscale}) with $L^{1/\nu} T$ constant and $\nu = 1.8$, the
average of our estimates from $g$ and $\chiSG$.
The value of the chaos exponent is $\zeta = 1.0$.
}
\label{plot:DJ}
\end{figure}

A scaling plot for chaos with $\D J$, following Eq.~(\ref{rJscale}) with
$L^{1/\nu}T$ constant (and $\nu = 1.8$),
is shown in
Fig.~\ref{plot:DJ}, for sizes between 4 and 10 with $\zeta=1.0$. 
The perturbation, $\D J$, lies in the range
$0.05 - 0.3$.
Trying different values of $\zeta$ we estimate 
\be
\zeta \ = \ 1.0 \pm 0.1 \quad \left( {\rm chaos} \ {\rm with} \ \D J \right) .
\label{zetaJ}
\ee
%We also did the same scaling analysis for data
%at $T=0.8$ and found the same chaos exponent within statistical
%uncertainties.

\begin{figure}
\epsfxsize=\columnwidth\epsfbox{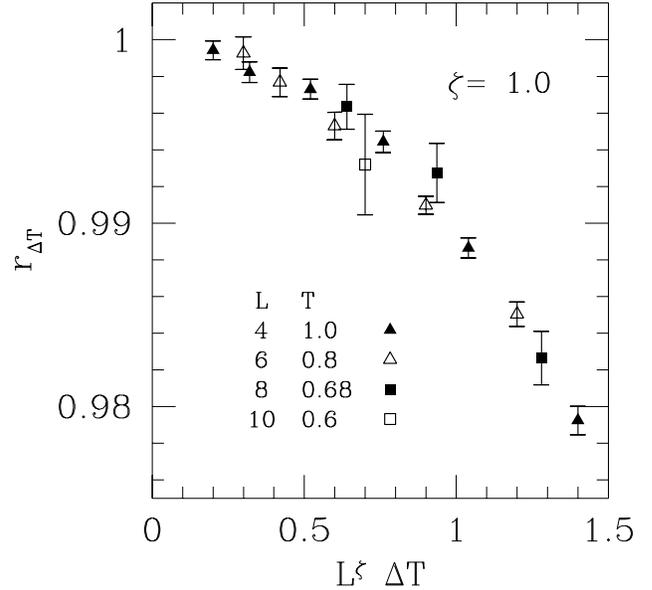}
\caption{
A scaling plot of $r_{\scriptscriptstyle \D T}$ according to
Eq.~(\protect\ref{rTscale})
with $L^{1/\nu} T$ constant and $\nu = 1.8$, the
average of our estimates from $g$ and $\chiSG$.
The value of the chaos exponent is $\zeta = 1.0$.
}
\label{plot:DT}
\end{figure}

A scaling plot for chaos with $\D T$, following Eq.~(\ref{rTscale})
with $L^{1/\nu}T$ constant (and $\nu = 1.8$),
is shown in
Fig.~\ref{plot:DT}, for $L \le  10$, and $\zeta=1.0$. 
The perturbation, $\D T$, lies in the range
$0.05 - 0.4$.
Trying different values of $\zeta$ we estimate 
\be
\zeta \ = \ 1.0 \pm 0.2 \quad  \left( {\rm chaos \ with} \ \D T  \right) \ .
\label{zetaT}
\ee
%We also did similar calculations at a somewhat lower temperature, $T=0.6$, but
%there
%the precision and the number of results
%are greatly limited by the lowest temperature $T-\D T/2$ 
%we can study in a reasonable computer
%time.
%and also by the small value of $\xi(T)$ when $T+\D T/2$
%is large which puts us out of the critical region.
%In any case,
%the chaos exponent is
%consistent with that found at $T=0.8$.

Note that the exponents for chaos with $\D J$ and $\D T$, given in
Eqs.~(\ref{zetaJ}) and (\ref{zetaT}), are equal within the uncertainties.  
We also see from Figs.~\ref{plot:DJ} and \ref{plot:DT} that the data for
$\rDT$ does not deviate very much from unity, as compared with the
data for $\rDJ$. This indicates that the {\em amplitude}
of chaos with $\D T$ is
smaller than that with $\D J$, even though the exponents are equal.

\section{Discussion}
\label{interpret}
Studying the two-dimensional Ising spin glass by finite temperature Monte Carlo
simulations, we are able to detect chaos with respect to both $\D J$ and $\D T$
and show that the chaos exponents are equal, as expected\cite{nh}.

We should point out that in order to define chaos with $\D T$ in the critical
region it is necessary that $\nu > 1 /\zeta$.
 To see this note that we 
need $l_{\Delta T}$ in Eq.~(\ref{zeta}) to be less that the correlation length,
$\xi \sim T^{-\nu}$, and also $\D T \ll T$ to be in the scaling region for
chaos. In our case, this
inequality is satisfied
(note that chaos in the critical region is also present when $T_c > 0$ 
\cite{nh,th}).
%also possible that we have an equality $\nu = 1 / \zeta$.
Chaos with $\D J$, on the other hand, can be defined
irrespective of the relative values of $\nu$ and $1/\zeta$.

Our estimates of the exponent, given in Eqs.~(\ref{zetaJ}) and (\ref{zetaT}),
%agrees with that obtained from the Migdal Kadanoff approximation\cite{nh},
%i.e. $\zeta = 0.74$.
%However, using
are consistent with the value $\zeta=0.95\pm0.05$ \cite{rsb} found from
exact ground state determinations.
A similar value was also found earlier by Bray and Moore\cite{bm}.
%There seem to be two possible reasons for this discrepancy.
%The first is that there
%may be systematic effects, not included our estimate of the error bar
%(which only allows for statistical uncertainties), which lower our estimate of
%$\zeta$. One possible source of such a correction is that the 
%bulk correlation length is comparable to (rather than much greater than, as
%is should be,)
%the largest lattice sizes used. Hence it is possible that we see a value for
%$\zeta$ in between that of the $T=0$ critical point, and that of the infinite
%temperature fixed point, where Migdal Kadanoff calculation\cite{nh} 
%gives $\zeta=0.5$.

%Another possibility is that the simple scaling picture, which relates exponents
%at finite temperature to those at $T=0$, may be invalid.

Finally we note that our value for $\nu$ agrees 
with work of Liang\cite{liang} who gets $\nu
\approx 2$ from Monte Carlo simulations. However
a much larger value is inferred  at $T=0$, from domain wall renormalization 
group
calculations\cite{hm}, i.e. $\nu = 4.2 \pm 0.5$, and from
exact ground state calculation\cite{rsb}, i.e. $\nu = 3.559 \pm 0.025$.
These discrepancies suggest a violation of the scaling picture of the spin
glass transition. 
Kawashima et al.\cite{khs} also find a
discrepancies in the scaling theory. 
%Our results are certainly incompatible with the scaling theory. To see this
%note that
%the scaling theory predicts $2(y + \zeta) = d_s$ where $y$ is the
%stiffness exponent for the domain walls (equal to $-1/\nu$ according to the
%scaling theory) and $d_s$ is the fractal dimension of the domain walls,
%which should be between 1 and 2.
%Inserting our values gives $d_s = 0.3 \pm 0.4$ in clear disagreement with the
%scaling picture. The fact that $d_s$ cannot be a fractal 
%surface dimension also appears at the critical point 
%\cite{th} where chaos  can be interpreted
%within Migdal Kadanoff renormalisation scheme.
If there are violations of the scaling picture in two dimensions
it would be very valuable to understand them since
similar violations may also occur in
higher dimensions with a finite $T_c$, and also
perhaps help resolve disagreements
between the droplet and mean field pictures. 

\acknowledgments
We thank Henk Hilhorst, Michel Gingras and David Huse for their comments
on the first version of this paper. We appreciate Christophe Ney's
help with C programming and thank
Thierry Biben for sharing his computer.
This research was supported by NSF Grant DMR 94--11964.
The work of MN was supported by a NATO fellowship and 
by the Centre National de la Recherche Scientifique.

\end{document}